\documentclass[prd,twocolumn,superscriptaddress,showpacs,floatfix,longbibliography,nofootinbib]{revtex4-1}
\usepackage{graphicx,amsfonts,amssymb,amsmath,xspace,physics}
\usepackage[utf8]{inputenc}
\usepackage[pdftex,breaklinks=true,colorlinks=true]{hyperref}
\usepackage{epstopdf}
\usepackage{color}
\definecolor{g}{rgb}{.1,0.4,.1} 
\definecolor{b}{rgb}{0,0.2,1}
\definecolor{rouge}{rgb}{0.82,0.,0.}
\definecolor{vert}{rgb}{0.,0.82,0.}
\definecolor{orange}{rgb}{1,0.5,0.}
\definecolor{bleu}{rgb}{0.,0.,0.82}
\definecolor{m}{rgb}{0.82,0.,0.82}
\definecolor{vert2}{rgb}{0.,0.5,0.}
\definecolor{rougeclair}{rgb}{1.0,0.7,0.7}

\newcommand{\beq}{\begin{equation}}
\newcommand{\be}{\begin{equation}}
\newcommand{\beqn}{\begin{eqnarray}}
\newcommand{\eeq}{\end{equation}}
\newcommand{\ee}{\end{equation}}
\newcommand{\eeqn}{\end{eqnarray}}

\newcommand{\bem}{\begin{pmatrix}}
\newcommand{\eem}{\end{pmatrix}}

\newcommand{\im}{\textrm{i}}

\newlength{\ldag}
\begin{document}

\title{Tunable Aharonov-Bohm-like cages for quantum walks}

\author{Hugo Perrin}
\email{perrin@lptmc.jussieu.fr}
\affiliation{Sorbonne Universit\'e, CNRS, Laboratoire de Physique Th\'eorique de la Mati\`ere Condens\'ee, LPTMC, 75005 Paris, France}

\author{Jean-No\"el Fuchs}
\email{fuchs@lptmc.jussieu.fr}
\affiliation{Sorbonne Universit\'e, CNRS, Laboratoire de Physique Th\'eorique de la Mati\`ere Condens\'ee, LPTMC, 75005 Paris, France}

\author{R\'emy Mosseri}
\email{remy.mosseri@upmc.fr}
\affiliation{Sorbonne Universit\'e, CNRS, Laboratoire de Physique Th\'eorique de la Mati\`ere Condens\'ee, LPTMC, 75005 Paris, France}

\date{\today}

\begin{abstract}
Aharonov-Bohm cages correspond to an extreme confinement for two-dimensional tight-binding electrons in a transverse magnetic field. When the dimensionless magnetic flux per plaquette $f$ equals a critical value $f_c=1/2$, a destructive interference forbids the particle to diffuse away from a small cluster. The corresponding energy levels pinch into a set of highly degenerate discrete levels as $f\to f_c$. We show here that cages also occur  for discrete-time quantum walks on either the diamond chain or the $\mathcal{T}_3$ tiling but require specific coin operators. The corresponding quasi-energies versus $f$ result in a Floquet-Hofstadter butterfly displaying pinching near a critical flux $f_c$ and that may be tuned away from 1/2. The spatial extension of the associated cages can also be engineered.
\end{abstract}


\maketitle

\section{Introduction } 

Two-dimensional (2d) electronic systems in a perpendicular magnetic field have been of special interest in condensed matter physics. They lead to various subtle effects such as the measured integer~\cite{Klitzing80} and fractional~\cite{Tsui82} quantum Hall states or the predicted self-similar Hofstadter butterfly~\cite{Hofstadter76} describing the structure of energy levels for tight-binding (TB) electrons versus the magnetic flux. 
The possibility of generating ``artificial" magnetic fields acting on cold atoms assemblies opens the way to different types of experiments in that direction \cite{Bloch12}.

About 20 years ago, an extreme localization effect was proposed, which occurs for TB models in certain periodic lattices, such as the 2d $\mathcal{T}_3$ (or dice) rhombus tiling~\cite{Vidal98} or the diamond chain (DC)~\cite{Vidal2000}, at half a magnetic flux quantum per plaquette. These so-called Aharonov-Bohm (AB) cages are due to a complete destructive interference preventing the particle to escape from finite clusters. Generic periodic structures, like the infinite square lattice studied in~\cite{Hofstadter76}, lead to energy bands for rational fluxes (measured in units of the flux quantum), and more complex density of states for irrational fluxes. In contrast, the $\mathcal{T}_3$ butterfly displays the surprising feature that the density of states pinches near a half flux quantum, leading to an energy spectrum consisting of three highly degenerate energy levels. The one at zero energy is present at any flux and is due to a chiral (bipartite) symmetry \cite{Sutherland86}. The other two are the result of destructive interferences tuned by the magnetic field. It was demonstrated that a particle, placed at initial time on one of the three different sites of the lattice, displays a quantum diffusion limited to a small cluster of sites, and eventually  periodically bounces back and forth to its original position. This effect disappears if the flux is tuned away from half a flux quantum~\cite{Vidal98} or if interactions between particles~\cite{Vidal2000} or disorder~\cite{Vidal2001} are introduced.

This predicted phenomenon inspired several experimental implementations or proposals, among which superconducting wire networks~\cite{Abilio99}, Josephson junction arrays~\cite{Pop2008}, cold atomic gases~\cite{moller12}, photonic lattices~\cite{Mukherjee18}, ions micro traps~\cite{Porras2011}, etc...

Here, we show that a related caging phenomenon can be obtained in the a priori different case of unitary quantum walks (QW). It displays several novel features that are absent from the original TB model~\cite{Vidal98}, such as the tuning of the critical flux and that of the cage size.

\medskip

The article is organized as follows. In Section~\ref{sec:mqw}, we introduce 1d and 2d discrete-time quantum walks in the presence of a perpendicular magnetic field. Then, in Section~\ref{sec:dc}, we study AB cages in the case of the 1d QW on the diamond chain. Next, in Section~\ref{sec:t3}, we obtain AB cages for 2d QW on the $\mathcal{T}_3$ lattice. Finally, in Section~\ref{sec:ccl}, we conclude with a summary and perspectives. Several Appendices provide details of calculations and further examples of AB cages, such as one leading to the tuning of the cage's spatial extension (see Appendix~\ref{ap:sizecage}).

\section{Magnetic Quantum walks}
\label{sec:mqw}
Quantum walks can be seen as a quantum generalization of classical random walks (see e.g. Refs.~\cite{Aharonov93,Meyer96} and~\cite{Kempe03} for review). Their main interest arises in the quantum information framework. We focus on discrete-time QWs. A particle, equipped with a finite dimensional internal state (with Hilbert ``coin'' space $\mathcal{H}_c$), is subject to unitary shifts along edges on a graph $\Lambda$ (with Hilbert space $\mathcal{H}_\Lambda=\mathrm{span}\lbrace\ket{i},i\in\Lambda\rbrace$), in a direction guided by the internal state. At each time step, a unitary coin operator $C$ is uniformly applied on the internal states, eventually affecting the following unitary shift $S$ and leading to the walk operator $W=S\, C$. After $N$ steps, the initial state $\ket{\psi(0)}$ becomes $\ket{\psi(N)}=W^N \ket{\psi(0)}$. QW protocols have already led to several experimental realizations in 1d with trapped atoms~\cite{karski09}, single photons~\cite{broome10} or Bose-Einstein condensates  in momentum space~\cite{dadras18}. In 2d, a generalization has been done with light~\cite{derrico18}, and the tentative addition of a magnetic field has also been  discussed~\cite{Yalcinkaya15, Arnault16, alberti19,Cedzich19}.

\begin{figure}[t]
\includegraphics[width=0.5\textwidth]{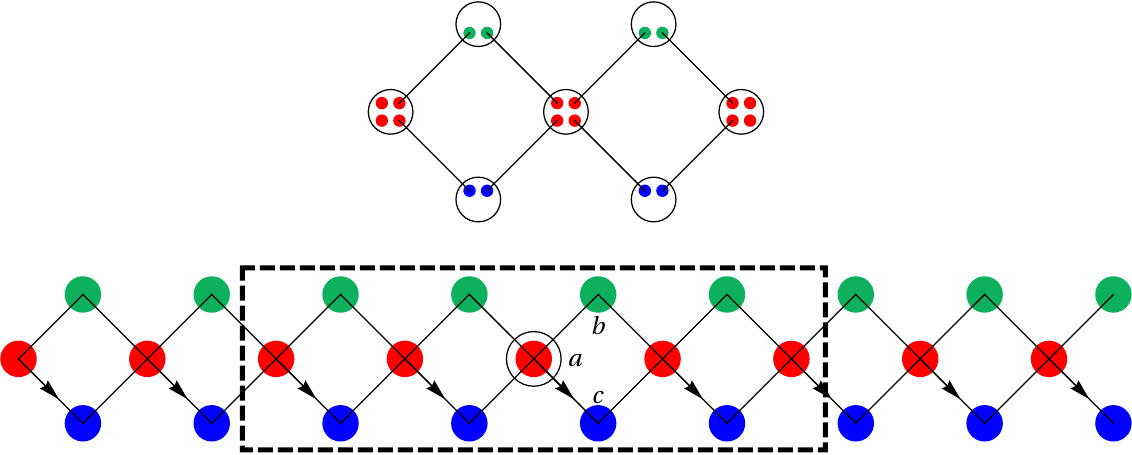}
\caption{Bottom: a piece of a diamond chain with hub sites $a$ (in red) and rim sites $b$ (in green) and $c$ (in blue). An arrow means a phase $e^{i 2\pi f}$, which is our gauge choice. The dashed rectangle indicates the maximal extension of a cage at critical flux, for an initial state localized on the circled $a$ site. Top: the QW Hilbert space is schematized with four (resp. two) basis states for $a$ (resp. $b,c$) sites, shown here as circles. Coins operate on states inside a circle, and shifts along  edges.}
\label{fig:dc}
\end{figure}

\begin{figure}[h]
    \centering
    \includegraphics[width=0.45\textwidth]{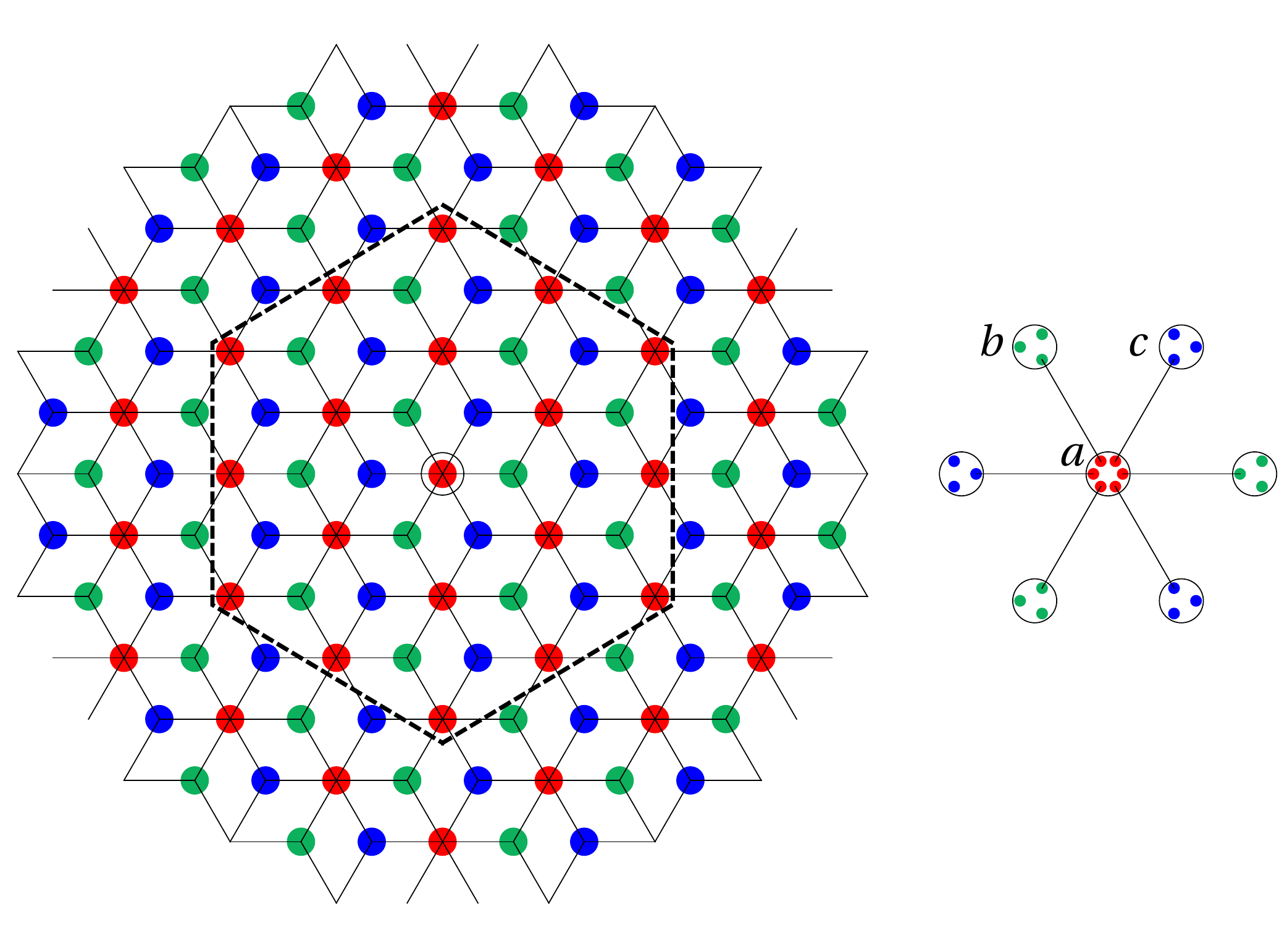}
    \caption{Left: a piece of $\mathcal{T}_3$ tiling with hub sites $a$ (in red) and rim sites $b$ (in green) and $c$ (in blue). The dashed hexagon indicates the maximal extension of a cage at critical flux centered on the initial circled $a$ site. Right: the QW Hilbert space is schematized, with six (resp. three) basis states for $a$ (resp. $b,c$) sites, shown here as circles.}
    \label{fig:T3lattice}
\end{figure}

We consider a QW in a perpendicular magnetic field and study cage effects on two periodic graphs: (i) the DC with four-fold coordinated ``hub'' sites $a$ and two-fold ``rim'' sites $(b,c)$ (Fig.~\ref{fig:dc}) (ii) the $\mathcal{T}_3$ tiling with six-fold hub sites $a$ and three-fold rim sites $(b,c)$ (Fig.~\ref{fig:T3lattice}). In each unit cell, the internal space size is $8$ for DC and $12$ for  $\mathcal{T}_3$. In a standard QW, the internal space codes for the direction of the walk. E.g., in 1d, a spin $1/2$ internal state is used to specify along which direction, left or right, to move. Applied to cases with sites of unequal connectivity would lead to internal spaces of different dimension on the different types of sites. Following Ref.~\cite{Ambainis2003} (and references therein), it is simpler to consider that a Hilbert space basis state is associated with a pair composed of a vertex and the directed edge incident on that vertex (see top of Fig.~\ref{fig:dc} and right of Fig.~\ref{fig:T3lattice}). The shift operator $S$ is then a sum of hopping terms between the two states associated with each edge (one per opposite sites of this edge) and is therefore unitary and hermitian. The operator $C_s$ shuffles the states associated with site $s=\{a,b,c\}$ and its different incident edges. The coin operator $C$ is the direct sum of $C_a,C_b$ and $C_c$. It may be hermitian, but generically does not commute with $S$, leading to a non trivial walk operator $W$. 
 
 We  define four types of $n\cross n$ coins associated with sites of coordination number $n=2,3,4,6$. Previous works have focused mainly on three types of coins~\cite{Kempe03}: Hadamard $H_n$, defined for sizes $n=2^m$, Grover $G_n$ and discrete Fourier transform $D_n$ coins, which can be constructed for any matrix size $n$. As $D_n$'s were found not to lead to caging effects, we shall not discuss them further. Grover coins read $G_n =(2/n) \,\mathbf{1}_n- \mathbb{I}_n$, where $\mathbf{1}_n$ is an $n\times n$ matrix full of $1$ and $\mathbb{I}_n$ is the identity matrix. 
 
For the two-fold sites, we use generic unitary  coins:
$$
U_2(\theta,\varphi,\omega,\beta)=
\begin{pmatrix}
\cos \theta \, {\rm e}^{\im  \beta}& -\sin \theta \, {\rm e}^{\im (\varphi+\omega)} \\
\sin \theta \, {\rm e}^{-\im  \omega} &\cos\theta \, {\rm e}^{i(\varphi-\beta)}
\end{pmatrix}.
$$
The standard Hadamard coin $H_2$ is $ U_2(\pi/4, \pi,0,0)$ and $SO(2)$ rotations correspond to $ U_2(\theta,0,0,0)$. For $n=3$, we use  $SO(3)$ rotations $R_3(\alpha,\gamma)$, of angle $\alpha$ around unit 3d vectors $\Vec{v}= \left( \frac{\cos\gamma}{\sqrt{2}},\sin{\gamma},\frac{\cos\gamma}{\sqrt{2}}\right) $.  Such a matrix reduces to $G_3$ whenever $(\alpha,\gamma)=(\pi,\sin^{-1}(1/\sqrt{3}))$. 
 For $n=4$, we use either  $H_4=H_2\otimes H_2$, or $G_4$, and finally $G_6$ is used for $n=6$. The basis conventions are detailed in Appendix \ref{ap:basis}. The shift operator reads
 $$S=\sum_{\left\langle( i,j),(i',j')\right\rangle}\ket{i,j}\bra{i',j'} +\text{h.c.},$$  
 where $i$ and $i'$ are neighbouring sites, and $j$ and $j'$ two opposite directed edges between  these two sites. The magnetic field \mbox{$B=|\boldsymbol{\nabla}\times \boldsymbol{A}|$} enters via a Peierls substitution~\cite{Peierls33}, i.e. the hopping terms in the shift $S$ get multiplied by a phase factor 
 $
{\rm e}^{\im \frac{2\pi}{\phi_0} \int_{i}^{i'} \textbf{dl}\cdot \boldsymbol{A}}
$, where $\boldsymbol{A}$ is the vector potential and $\phi_0=h/e$ the flux quantum~\cite{Yalcinkaya15, alberti19, Cedzich19}. 
Being unitary, $W$ can be written as the exponential ${\rm e}^{- \im H_\text{eff}}$ of an effective hamiltonian $H_\text{eff}$, whose evolution is only considered at integer times. Its eigenvalues are pure phases, called quasi-energies and defined modulo $2\pi$.

\section{AB cages on the diamond chain} 
\label{sec:dc}
In contrast to 2d tilings, for the DC (see Fig.~\ref{fig:dc}), it is possible to find a gauge that preserves the periodicity of the lattice (i.e. a unit cell containing 8 basis states). We choose it as a phase ${\rm e}^{\im 2\pi f}$ on one of the four edges (see bottom of Fig.~\ref{fig:dc}), with the reduced flux $f$ defined as the magnetic flux per plaquette in units of $\phi_0$, i.e. $f=\phi_0^{-1} \oint_\text{pl.} \textbf{dl}\cdot \boldsymbol{A}$. Using translation invariance, it is possible to diagonalize the $W$ operator into $8\times8 $ $k$-dependent blocks $W(k)$ with vanishing diagonal $4\times 4$ sub-blocks, where $k$ is a wave-vector inside the first Brillouin zone. Then, $W^2$ is made of $4\times 4$ (isospectral, see Appendix \ref{ap:isospectrum}) diagonal blocks, with eigenvalues ${\rm e}^{\im E(k)}$, so that $W$ eigenvalues are $ {\rm e}^{\im \epsilon(k)}$, with $\epsilon=E/2$ and $E/2+\pi$.

Different coins can be used on the $b$ and $c$ sites. For example, with $C_a=G_4$, $C_b=U_2(\theta,\varphi,0,\beta)$ and $C_c=U_2(\theta,\varphi,\omega,\beta)$, one gets the following ``quasi-energies" $\epsilon(k)$, leading to four $\beta$-independent bands $\epsilon(k)$ and four flat bands $\epsilon^{fb}$:
$$
\begin{aligned}
\epsilon(k)&=\frac{\varphi +\pi}{4} \pm \frac{1}{2}\cos ^{-1} \left[\sin \theta \cos \left(\pi f-\frac{\omega}{2}\right) \times \right. \\
 &\left.\cos \left(\pi f-\frac{\omega}{2}+k+\frac{\pi -\varphi}{2} \right) \right] \pm \frac{\pi}{2}  \\
\epsilon^{fb}&=\pm \frac{\pi}{2}+\frac{\varphi}{4}\pm \frac{1}{2} \cos^{-1}(\cos(\beta-\varphi/2)\cos\theta)  .
\end{aligned}
$$

\begin{figure}[!!h]
 \includegraphics[width=0.5 \textwidth]{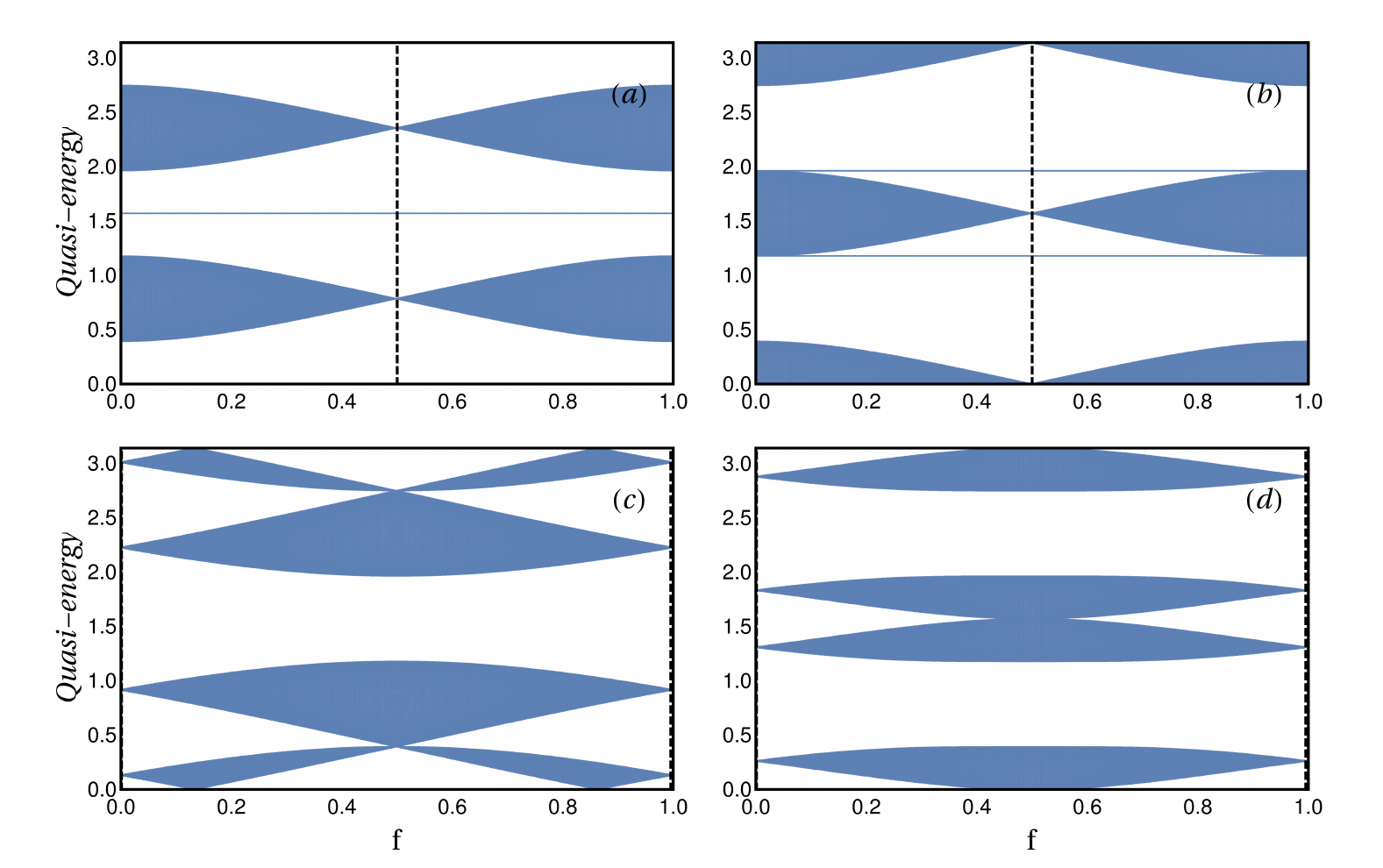}
\caption{Diamond chain quasi-energies spectra, in the range $[0,\pi]$ (repeated under a $\pi$ translation), plotted versus the reduced flux $f$ (invariant under translation of period $1$), for the coins $\{C_a,C_b=C_c\}$ equal to: (a) $\left\lbrace G_4 \, , U_2(\pi/4,\pi,0,0)\right\rbrace $, (b)  $\left\lbrace G_4 \, , U_2(\pi/4,0,0,0)\right\rbrace $, (c) $\left\lbrace H_4 \, , U_2(\pi/4,\pi,0,\pi)\right\rbrace$, (d) $\left\lbrace H_4 \, , U_2(\pi/4,0,0,0)\right\rbrace$. Varying $\omega$ would lead to an horizontal shift of the whole pattern. The critical flux [either 1/2 for (a) and (b) or $0\sim1$ for (c) and (d)] is indicated in each case by a dashed vertical line.}
\label{fig:butterflychaineboucle}
\end{figure}

The quasi-energies versus flux patterns display symmetries: two translations, $f\longleftrightarrow 1+f$ (due to the Peierls substitution) and $\epsilon \longleftrightarrow \epsilon +\pi$ (due to a bi-partite graph~\cite{alberti19}) and two mirrors, $f \longleftrightarrow -f+\omega/\pi$ (with  $k \longleftrightarrow -k-\pi+\varphi$) and $\epsilon \longleftrightarrow -\epsilon +\varphi/2$ (with  $k \longleftrightarrow k+\pi$).  Fig.~\ref{fig:butterflychaineboucle}-a,b show the resulting $W$ quasi-energies displayed between $0$ and $\pi$ owing to the above translation symmetry, for $C_a=G_4$  and symmetric coins ($C_b=C_c$, meaning that $\omega =0$). They display two striking features: (i) the existence of flat bands (versus $k$ and $f$, which are even $\theta$-independent whenever $\beta-\varphi/2=\pm \pi/2$); (ii) a pinching of the dispersive bands. The critical value $f_c$ at which the pinching  occurs can be tuned at will, if $\omega\ne 0$ (therefore $C_b\neq C_c$), as $f_c=1/2+\omega/2\pi$, and even made to vanish (not shown in the Figure). 

These two features (i) and (ii) are independent: indeed, using a $H_4$ coin on $a$ sites (instead of $G_4$) still leads  to a pinching, but without flat bands (Fig.~\ref{fig:butterflychaineboucle}-$c,d$). Here $f_c=\omega/2\pi$ and pinching occurs at vanishing magnetic field whenever identical coins are applied on sites $b$ and $c$ (i.e. $\omega=0$). At $f_c$ the 8 quasi-energies read:
$$
\begin{aligned}
&\epsilon=\frac{\pi}{4} \pm\frac{\pi}{2}+\frac{\varphi}{4} \pm\frac{1}{2} \times \\
 &\cos^{-1}\left(\frac{\sin (\beta -\frac{\varphi}{2}) \cos\theta \pm \sqrt{2 \sin^2 \theta + \cos^2 \theta \sin^2(\beta-\frac{\varphi}{2})}}{2}\right)
\end{aligned} 
 $$
 
\begin{figure}[!ht] 
\centerline{
\includegraphics[trim={0 1.5cm 0 2cm},clip,width=0.5\textwidth]{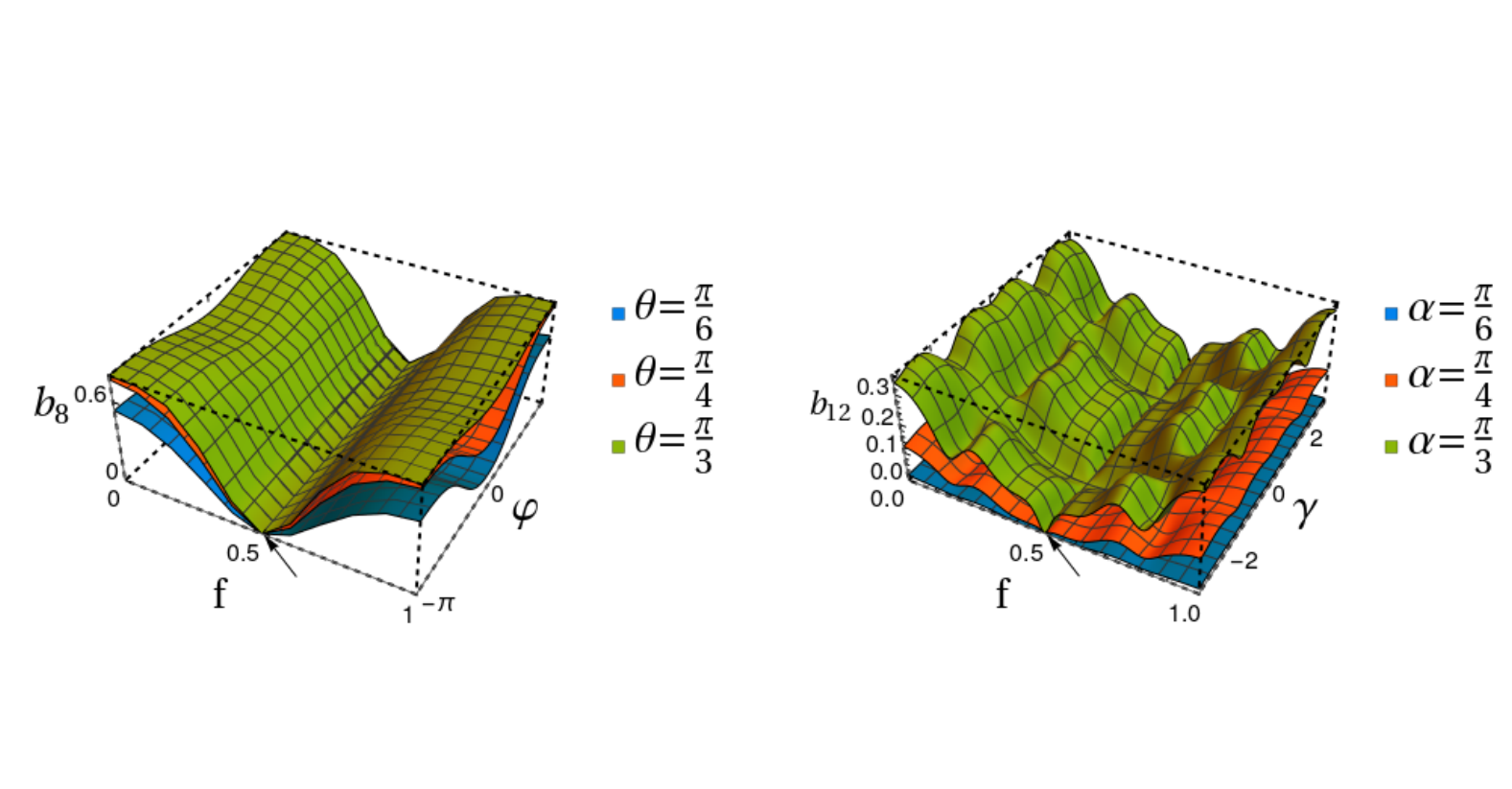} 
}
\caption{Arnoldi recursion coefficients: a) DC case with a $G_4$ coin: plot of $b_8$ versus $(f,\varphi)$ for $\beta=0$ and $\theta=\pi/3,\pi/4,\pi/6$; 
b) $  \mathcal{T}_3$ case: plot of  $b_{12}$ versus $(f,\gamma)$ for $\alpha=\pi/3,\pi/4,\pi/6$. The critical flux $f_c=1/2$ is indicated by an arrow.} 
 \label{arnoldi} 
 \end{figure}

The pinching is associated with an AB-like caging effect. In the TB case~\cite{Vidal98}, the AB caging was proved by analyzing the local density of states, with a Lanczos tridiagonalization showing a vanishing recursion coefficient for  $f_c=1/2$ and signing a non-propagating quantum evolution. Here we use a related approach adapted to unitary operators: $W$ is transformed into an almost triangular matrix with an added subdiagonal (Hessenberg form) using the Arnoldi iteration (see Appendix \ref{ap:Arnoldi} and Ref. \cite{Arnoldi1951}). The latter consists in starting from a given state $\ket{0}$ and then iteratively applying $W$. Each obtained new state $|n+1\rangle$ is made orthogonal to all previous ones $\{|m\rangle, m\leq n\}$ according to:
$$
b_{n+1}\ket{n+1}=W\ket{n}-\sum_{m\leqslant n}\bra{m}W\ket{n}\ket{m}.
$$ 
 A vanishing  coefficient $b_{n}$ with $n=n_c$ terminates the iteration and indicates a caging effect. Indeed, starting from a localized state, this proves that only a finite fraction of states can be reached, leading to an evolution inside the subspace spanned by the states $\{\ket{n}, n \leq n_c\}$.
A vanishing coefficient $b_8$ is indeed found for all above described DC cases, with an initial state localized at any internal state on an $a$ site, see Fig.~\ref{arnoldi}-a. This leads to a cage of maximal radius twice the size of the unit cell (see Fig.~\ref{fig:dc}).

A last point concerns the time evolution  inside a cage. The above quasi-energies are incommensurate for generic $(\theta, \varphi,\beta) $ values, which leads to a quasiperiodic time behavior. A periodic behavior can nevertheless arise as displayed in Table ~\ref{Table}. A sketch of the dynamics of a period 8 cage is shown in Appendix \ref{ap:movie}.

\begin{table}
\begin{tabular}{|c|c|c|c|c|c|}
\hline
Graph & $C_a$ &  $C_c$ & FB &  $f_c$ & Period\tabularnewline\hline\hline
DC & $G_4$ & $U_2(\theta,\varphi,\omega,\beta)$ generic & Yes &$\frac{1}{2}+\frac{\omega}{2\pi}$ & QP\tabularnewline\hline
DC & $G_4$ & $U_2(\theta,\varphi,\omega,\frac{\pm\pi+\varphi}{2})$ & Yes  &$\frac{1}{2}+\frac{\omega}{2\pi}$ & $8$\tabularnewline\hline
DC & $G_4$ & $U_2(\frac{2\pi p}{q},\varphi,\omega,-\frac{\varphi}{2})$ & Yes &$\frac{1}{2}+\frac{\omega}{2\pi}$ & LCM$(4,2q)$\tabularnewline\hline
DC & $H_4$ &  $U_2(\theta,\varphi,\omega,\beta)$ generic  & No & $\frac{\omega}{2 \pi}$ &QP\tabularnewline\hline
DC & $H_4$ & $U_2(\frac{\pi}{4},\pi,\omega,0)$ & No & $\frac{\omega}{2\pi}$ & $24$\tabularnewline\hline
DC & $H_4$ & $U_2(\frac{\pi}{4},\frac{\pi}{2},\omega,0)$ & No & $\frac{\omega}{2\pi}$ & $10$ \tabularnewline\hline
DC & $H_4$ & $U_2(\frac{\pi}{4},0,\omega,0)$ & No & $\frac{\omega}{2\pi}$ & $12$ \tabularnewline\hline
$\mathcal{T}_3$& $G_6$ & $R_3(\alpha,\gamma)$ generic & Yes & $\frac{1}{2}$ &QP\tabularnewline\hline
$\mathcal{T}_3$& $G_6$ & $\tilde{R_3}(\alpha,\gamma,\frac{-2\pi}{3})$ & Yes & $\frac{1}{6}$ &QP\tabularnewline\hline
$\mathcal{T}_3$& $G_6$ & $R_3(\frac{2\pi}{3} ,\gamma)$ & Yes & $\frac{1}{2}$ &12\tabularnewline\hline

\end{tabular}
\caption{Examples of caging effects for QW on DC and $\mathcal{T}_3$ lattice for different coins $C_a$ and $C_c$ [with $C_b=C_c(\omega=0)$]. The table indicates the critical flux $f_c$ for the spectral pinching, FB tells whether flat bands are present, and the last column  specifies the period of the time evolution when it exists (QP referring to a quasiperiodic case). }
\label{Table}
\end{table}

\section{AB cages on the $\mathcal{T}_3$ lattice }
\label{sec:t3}
\begin{figure}[h]
\begin{center}
   \includegraphics[width=0.5\textwidth]{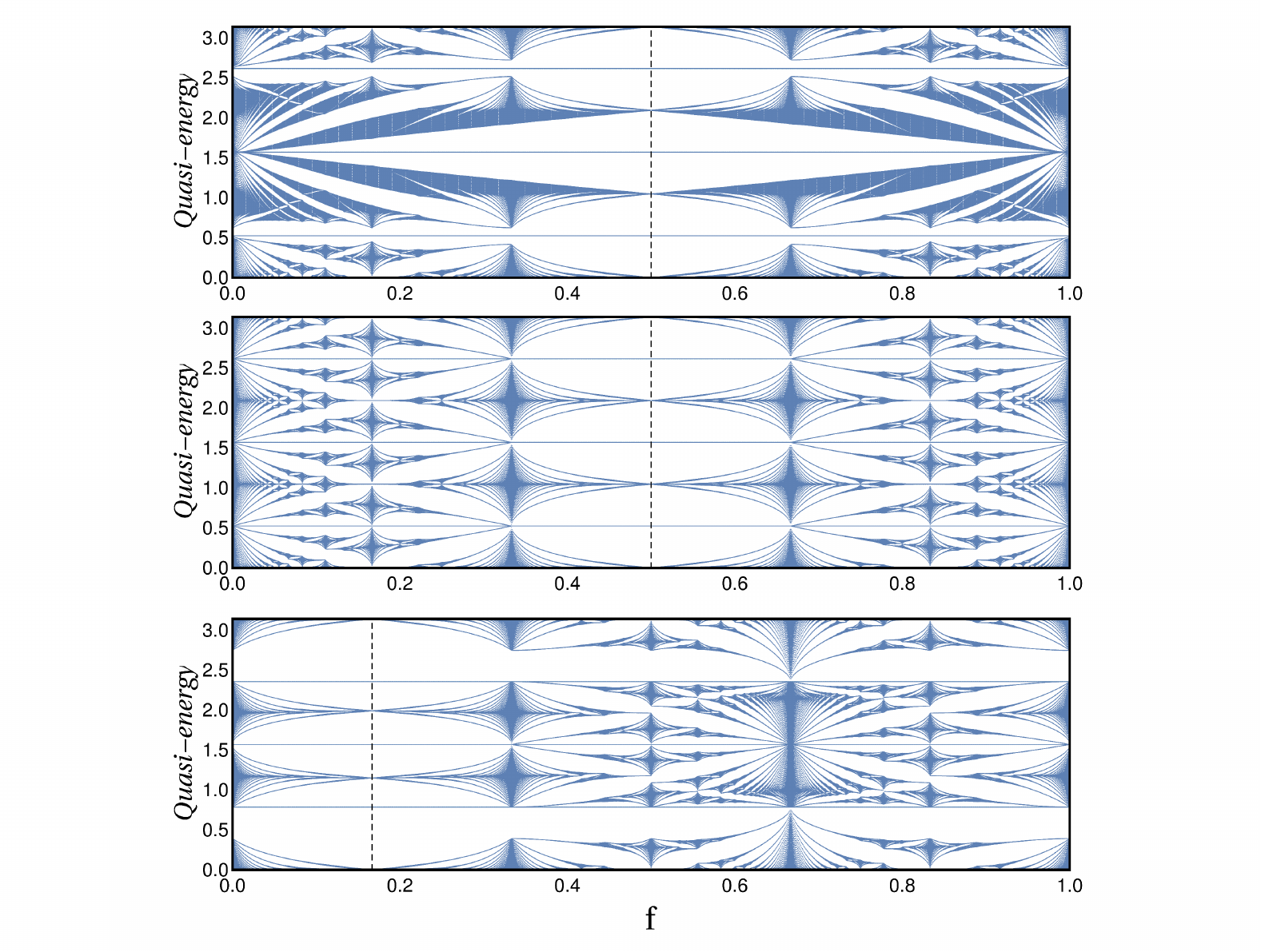}
  \caption{Floquet-Hofstadter butterfly on the $\mathcal{T}_3$ lattice (i.e. quasi-energy spectra of a QW versus magnetic flux $f$) for three values of the $R_3$ coin operator. Top: $\alpha=2\pi/3, \gamma=\sin^{-1}(1/\sqrt{3}), \omega=0$, Middle: $\alpha=2\pi/3, \gamma=0, \omega=0$, Bottom: $\alpha=\pi/2, \gamma=\sin^{-1}(1/\sqrt{3}), \omega=-2\pi/3 $. The vertical dashed lines indicate the critical flux $f_c=1/2$ (top), 1/2 (middle) and 1/6 (bottom).}
    \label{fig:spectreT3}
 \end{center}
\end{figure}

Next, we consider a 2d QW on the $\mathcal{T}_3$ lattice (see Fig.~\ref{fig:T3lattice}). Using the $G_6$ coin on $a$ sites, we found suitable $R_3(\alpha,\gamma)$ coins for $(b,c)$ sites that lead to caging.  As in the DC case, we allow for a different $R_3$ coin operating on the $c$ and $b$ sites by introducing an angle $\omega$ entering
$\tilde{R_3}(\alpha,\gamma,\omega)_{i,j}=R_3(\alpha,\gamma)_{i,j} \, e^{-i\omega\sum_{k}\varepsilon_{ijk}}$
with $\varepsilon_{ijk}$ the skew-symmetric Levi-Civita tensor. This unitary coin is no longer a 3d rotation. As in the DC, the introduction of $\omega \neq 0$ breaks time-reversal symmetry. Using a Landau gauge, periodicity is present in the (say) $y$ direction (with a magnetic unit cell containing $12q$ states), and $W$ can be diagonalized for rational $f=p/q$ into $12q\times12q$ blocks,  leading to a Floquet-Hofstadter butterfly~\cite{Asboth17} shown in Fig.~\ref{fig:spectreT3}. It displays flat bands, apparent self-similar sub-patterns and  pseudo-Landau levels. Here, we focus on the spectral pinching that occurs near $f_c=1/2$. The quasi-energies, doubly degenerate and $k$-independent, read:
$$
\begin{aligned}
 \epsilon&=0,\pm\frac{\pi}{2},\pi,\frac{\pi}{2}\pm\frac{\alpha}{2},-\frac{\pi}{2}\pm\frac{\alpha}{2},\\
& \frac{\pi}{2}\pm\frac{1}{2}\cos^{-1}\left(\frac{2+\cos\alpha}{3}\right),
-\frac{\pi}{2}\pm\frac{1}{2}\cos^{-1}\left(\frac{2+\cos\alpha}{3}\right).
\end{aligned} 
 $$ 
 These expressions show that the dynamics of the QW cages can be periodic or not depending on whether quasi-energies differences are commensurate or not (see Appendix \ref{ap:dynamics}).


Using the Arnoldi iteration, one  numerically finds that the $b_{12}$ coefficient vanishes at $f_c=1/2$, as shown in Fig.~\ref{arnoldi}-b. The associated AB cage depends on the precise chosen initial state, the largest one, displayed in Fig.~\ref{fig:T3lattice}, being larger than in the TB case.

The flux values $\pm 1/3$ are particular: a gauge exists, which shares the tiling periodicity (see Appendix \ref{ap:jauge}). As a consequence, $W$ can be diagonalized into $12\times 12$ blocks $W(k_x,k_y)$. Asymmetric $\tilde{R_3}$ coins can then change the critical $f_c$ value: when $\omega=-2\pi/3$, $f_c$ is shifted from $1/2$ to $1/6$, according to $f_c=1/2+\omega/2\pi$. In contrast to the DC case, here $f_c$ cannot be tuned continuously.

\section{Conclusion}
\label{sec:ccl}
As in the TB case, the AB cages in a QW result from a destructive interference tuned by the magnetic field entering through  Peierls phases. However, in the present case, the action of the coin operator on the internal degrees of freedom is crucial for the caging to occur. Amplitudes vanish on those states that would otherwise allow the particle, upon further shift, to escape the cage. As a result, the AB cages in the QW case have a larger size than their TB counterparts.

Playing with different type of coins, cages can occur at critical fluxes others than $1/2$. Necessary conditions for this tuning of $f_c$ are the existence both of a gauge respecting the translation symmetries of the structure, and of a coin that breaks time-reversal symmetry (corresponding to $\omega\neq 0$ in the rim coin operators) and partially compensates the effect of the applied magnetic field. For example, in the DC, the $G_4$ (resp. $H_4$) coin creates AB cages at flux $f_c=1/2$ (resp. 0). 

In addition, the coin on hub sites may be seen as a kind of magnetically-tunable gate that behaves as a mirror at a critical flux $f_c$. Therefore, in DC, a periodic substitution by one type of coin (say $G_4$ in an array of $H_4$) allows one to engineer AB cages (here at $f_c=1/2$) of arbitrary size by tuning the $G_4$ coins inter-distance (see Appendix \ref{ap:sizecage}).

The QW AB cages and the corresponding Floquet-Hofstadter butterfly should be readily observable. Recently, a concrete experimental implementation of a magnetic QW in 2d has already been proposed~\cite{alberti19}. The extra challenge is to implement such a QW on a lattice with varying coordination number. QW on such lattices have been theoretically studied for many years~\cite{Ambainis2003}, but we are not aware of an efficient experimental realization to date. A brute force solution would be to have a large internal space (8 for DC or 12 for $\mathcal{T}_3$). Finally, a striking confirmation of the above predictions would be to test the tunability of the critical flux at which caging occurs and that of the cage size.
 
Although QW may be seen as a subclass of periodically-driven Hamiltonian systems, our proposal for the DC is significantly different from the AB cages seen in~\cite{Mukherjee18} that uses Floquet engineering to realize the TB model of~\cite{Vidal2000} and does not involve the action of coins in an internal space.

\acknowledgements
We thank Janos Asb\'oth for useful discussions.

\appendix
\section {Basis conventions}
\label{ap:basis}
The differents coin operators, acting at a given site, are given in the main text. We add here  our chosen basis ordering convention in the different cases, relative to the matrix representation of different coins.
\begin{figure}[h]
   \includegraphics[width=0.5\textwidth]{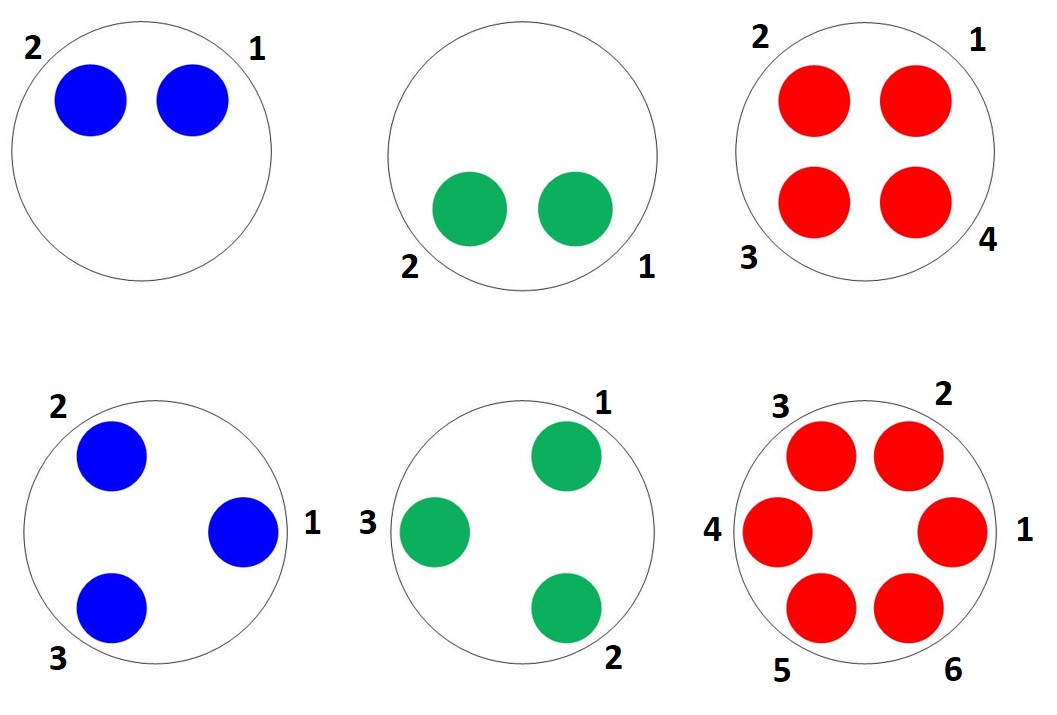}
    \caption{Basis states ordering convention for coordinated sites. (Top) diamond chain with two 2-fold (rim $b,c$) and one 4-fold (hub $a$) sites. (Bottom) $\mathcal{T}_3$ tiling with two 3-fold (rim $b,c$) and one 6-fold (hub $a$) sites. }
  
    \label{bases}
\end{figure}
 \par On the diamond chain, the right (resp. left) internal state of sites $b$ and $c$ corresponds to the first (resp. second) vector in the matrix representation: $\begin{pmatrix}1\\0
\end{pmatrix}$. The action of the coin is the multiplication by the generic unitary operator $U_2(\alpha,\varphi,\omega,\beta)$. For sites $a$, the first vector is the state on the right-up side of the site, and the basis is anti-clockwise oriented. For instance, a quantum walker localized on the left down internal state of a 4-fold site, indicated by the number 3 on top of Fig.~\ref{bases} is represented by the vector: $\begin{pmatrix}
0\\0\\1\\0
\end{pmatrix}$. Then, the application of the Grover or the Hadamard matrix defines the action of the coin.

\section {Isospectrality of $W^2$ sub-blocks }
\label{ap:isospectrum}

In the two examples treated here, the underlying graph is bipartite. The Hilbert space basis is the union of two parts $\mathcal{B}_a$ and $\mathcal{B}_{b,c}$, and the action of the unitary operator $W$ is bipartite: it sends components in $\mathcal{B}_a$ onto components in $\mathcal{B}_{b,c}$ and vice versa. As a consequence $W^2$ sends each sub-basis onto itself, and appears therefore block-diagonal. 

Now, suppose that $\ket{\psi}=\ket{\psi_a}+\ket{\psi_{b,c}}$ is an eigenvector of $W$, with eigenvalue $e^{\im \epsilon}$, with
 $\ket{\psi_a}$ (resp. $\ket{\psi_{b,c}}$) being the subpart of $\ket{\psi}$ in $\mathcal{B}_a$ (resp. $\mathcal{B}_{b,c}$).  $\ket{\psi}$ is an eigenvector of $W^2$, with eigenvalue $e^{2 \im \epsilon}$. In each diagonal sub-blocks of $W^2$, spanned by  $\mathcal{B}_a$ and $\mathcal{B}_{b,c}$, we therefore have that $\ket{\psi_a}$ and $\ket{\psi_{b,c}}$ are separately eigenvectors of $W^2$ with the same eigenvalue $e^{2 \im \epsilon}$, the two blocks being therefore isospectral.

\section {Arnoldi iteration}
\label{ap:Arnoldi}

The Arnoldi algorithm \citep{Arnoldi1951} amounts to tranform a generic square matrix into a so-called ``Hessenberg form", which is an almost triangular form. More precisely, it results to a (say upper) triangular form plus the lower sub-diagonal.

Here we apply the Arnoldi iteration to reduce the unitary QW operator. We start with a localized state $\ket{0}$ (the initial condition stating where the wavefunction is concentrated at $t=0$) and successively apply  the QW operator $W$. At each step, a new state is obtained from which the part which belongs to states already explored is removed: $$b_{n+1}\ket{n+1}=\hat{W}\ket{n}-\sum_{m\leqslant n}\bra{m} W\ket{n}\ket{m}.$$ As a consequence, a new orthonormal basis ${\ket{n}}$ is generated in which $\hat{W}$ has an Hessenberg form. Note that, when the operator is hermitian, the Arnoldi algorithm reduces to the Lanczos tridiagonalization.
 
The $b_n$ are defined positive, and the algorithm ends whenever a new generated state  is null ($b_j=0$). As a consequence, starting from the (localized) state $\ket{0}$, the quantum walk explores only a finite number of states, leading to a cage trapping.

\begin{widetext}
\section {Sketch of the cage effect in a simple DC case }
\label{ap:movie}
Fig.~\ref{sketch} shows a sketch of a periodic quantum walk on the diamond chain, at the critical value corresponding to a caging effect. We take the simple example of a period 8 quantum walk, with parameters given in the caption.
\begin{figure*}[h!]
   \includegraphics[width=\textwidth]{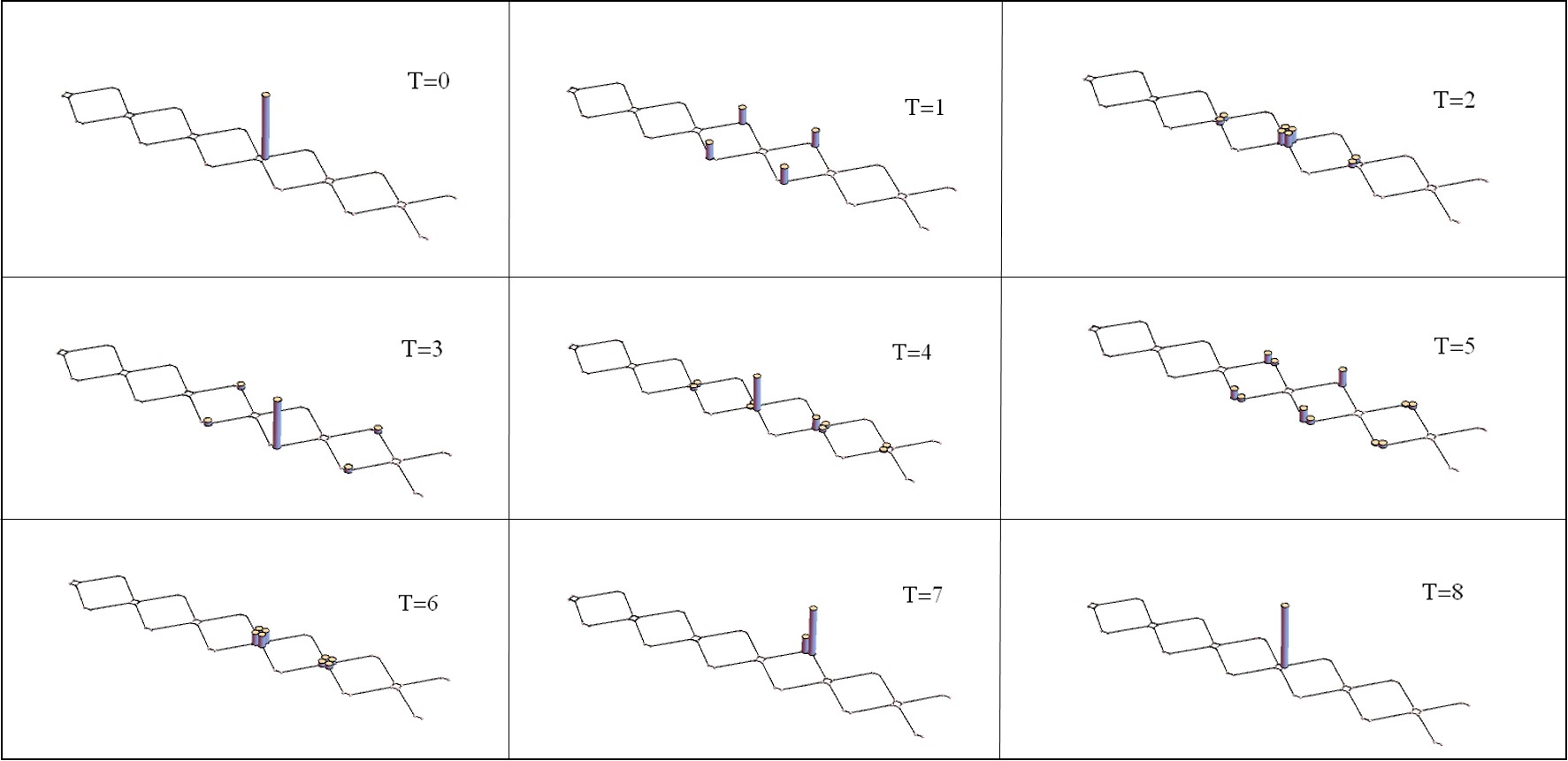}
  \caption{Sketch of the dynamics a period 8 quantum walk on the diamond chain with coins $C_a=G_4, C_b=C_c= U_2(\pi/4,0,0,-\pi/2)$ and $f_c$=1/2. The probability to find the walker on each eigenbasis state is shown as a function of integer time $T$.}
 \label{sketch}
\end{figure*}
\end{widetext}

\section {Criteria for periodic dynamics at the caging critical value $f_c$}
\label{ap:dynamics}

At the critical caging value $f_c$, an interesting question is to analyse  the confined dynamics. It is periodic whenever the differences between quasi-energies are commensurate. In the tight binding case the dynamics were found to be periodic in all cases (DC and $\mathcal{T}_3$). QW caging effects are found here with a larger set of parameters and are generically quasiperiodic. However, the analysis of the quasi-energies in the expressions given in the main text allows one to find some conditions for periodicity.

\subsection{DC case}
The fact that caging effects are found with a generic $U(2)$ coin $ U_2(\theta,\varphi,\omega,\beta)$ leads to a large spectrum of parameters. In the case $C_4=G_4$, one finds for instance that the dynamics is periodic whenever $ \cos^{-1}(\cos(\beta-\varphi/2)\cos\theta)$ is commensurate with $\pi$. For instance, if $\beta-\varphi/2=0$, we find a periodic dynamics, with period LCM$(4,2q)$, whenever $\theta=2 \pi p/q$ with $(p,q)\in\mathbb{Z}$. On the other hand, if $\beta-\varphi/2=\pm \pi/2$, the resulting dynamics has a period 8 for any $\theta$.

\subsection{ $\mathcal{T}_3$ case}
Substracting the quasi-energies at the critical $f_c$ value, we have that $\alpha$ must be commensurate with $\pi$, say $\alpha=\pi p_1/q_1$ and satisfy an additional more complicate equation:
 $\cos^{-1}\left((2+\cos\frac{\pi p_1}{q_1})/3\right)=\pi\frac{p_2}{q_2}$. If this is verified, the quantum walk is periodic, with period $4\times\mathrm{LCM}(q_1,q_2)$. We find that for $\alpha=2\pi/3$, the dynamics has period $12$, and numerically that no other solution exists for $q_1,q_2\leq 100$.

\section {Example of a periodic gauge at flux -1/3 for the $\mathcal{T}_3$ tiling.}
\label{ap:jauge}
\begin{figure}[h]
   \includegraphics[width=0.3\textwidth]{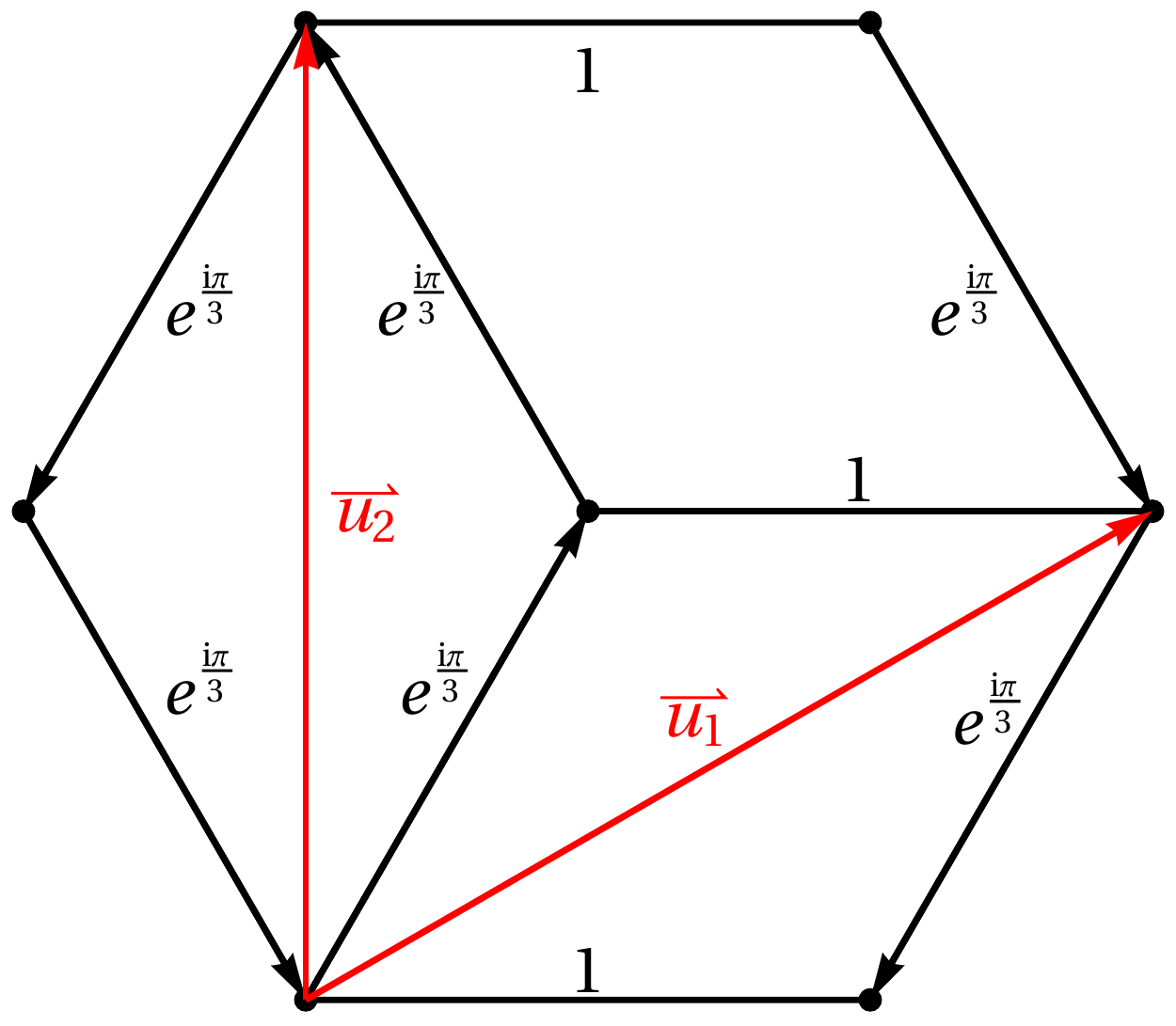}
  \caption{An example of a periodic gauge at flux $-1/3$ for the $\mathcal{T}_3$ tiling}
    \label{gauge}
\end{figure}

Fig.~\ref{gauge} shows an example of a gauge at flux -1/3 sharing the $\mathcal{T}_3$ tiling periodicity (whose unit vectors are displayed in red).

\section{Cages with tunable size in the DC case}
\label{ap:sizecage}
In the DC case, Grover $G_4$ and Hadamard $H_4$ coins on the $a$ hub sites create cages at different fluxes. Playing with this new feature allows one to control the extension of the cage. We briefly describe an example of that procedure. Start, at flux $1/2$, with a chain of $H_4$ coins on the hub sites, which do not therefore create cages (see Fig.~\ref{fig:butterflychaineboucle}), and substitute some  $H_4$ by $G_4$ coins in a periodic manner.   This induces AB cages, whose size depends both on the $G_4 - G_4$  distance (the superlattice period), and on the initial state of the quantum walk. 
 
 The rules to determine the extension of a cage  starting from an initial  state localized on a site $a$,  in the DC unit cell labelled $n_0$, is:
\begin{itemize}
\item First rule : The coin on the initial site $n_0$ is irrelevant for the cage to occur. What counts are coins applied on neighbouring sites.
\item Second rule : On the right-hand side, the QW spreads until it meets a substitution coin which will stop it on the right next site. 
\item Third rule : On the left-hand side, the coin on the first neighbour of the initial site does not matter. Then, the QW spreads and stops exactly when it meets a substitution coin.
\end{itemize}

\begin{figure}[h]
   \includegraphics[width=0.5\textwidth]{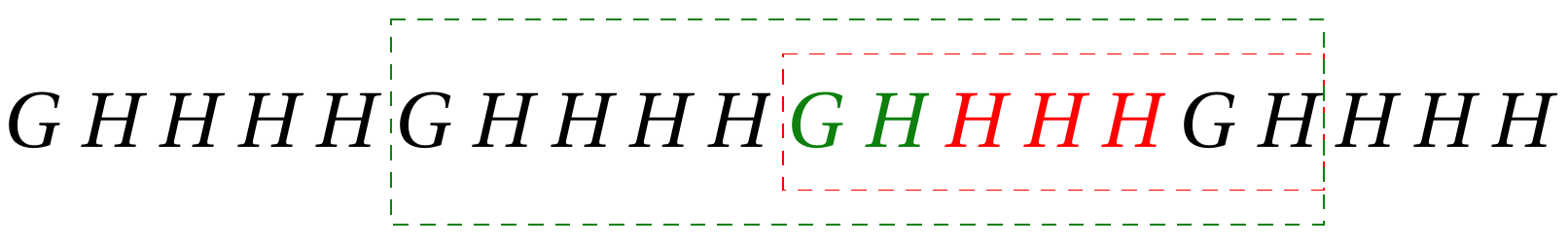}
  \caption{Diamond chain where only hub sites $a$ are represented by the letter $H$ (for Hadamard) or $G$ (for Grover) standing for the coin used on the hub site. In between hub sites, composite shift operators $K=S.U_2.S$ are always present but not represented. Dashed red (resp. green) box delimits the extension of the cage for an initial state on red (resp. green) sites at flux $1/2$.}
    \label{chaineHG}
\end{figure}

Fig.~\ref{chaineHG} gives an example, at flux $1/2$, for a $G_4$ coin inserted every 5 hub sites (in a chain containing only $H_4$ coins otherwise). Cages are represented by dashed boxes where the color indicates the location of the initial state. To understand the above rules, it is useful to operate a basis change within the internal space at the hub sites,  splitting between the right ($R$) and left ($L$) part, and combining the upper and lower part under the  form :
\begin{eqnarray*}
\ket{R^+}=\frac{1}{\sqrt{2}} \begin{pmatrix}1\\0\\0\\1\end{pmatrix},\, \ket{R^-}=\frac{1}{\sqrt{2}} \begin{pmatrix}1\\0\\0\\-1\end{pmatrix},\\
 \ket{L^+}=\frac{1}{\sqrt{2}} \begin{pmatrix}0\\1\\1\\0\end{pmatrix},\, \ket{L^-}=\frac{1}{\sqrt{2}}\begin{pmatrix}0\\1\\-1\\0\end{pmatrix}
\end{eqnarray*}
The effect of $H_4$ and $G_4$ coins onto these new basis states reads :

$$H_4:\ \left\{
    \begin{array}{ll}
      H_4\ket{R^+}\propto \ket{R^+}\\
      H_4\ket{R^-}\propto \ket{L^+} \\
      H_4\ket{L^+}\propto \ket{R^-}\\
      H_4\ket{L^-}\propto \ket{L^-} 
    \end{array}
\right. ,\ G_4:\ \left\{
    \begin{array}{ll}
      G_4\ket{R^+}\propto \ket{L^+}\\
     G_4\ket{R^-}\propto \ket{R^-} \\
      G_4\ket{L^+}\propto \ket{R^+}\\
     G_4\ket{L^-}\propto \ket{L^-} 
    \end{array}
    \right. $$
    
 These new basis states are therefore either eigenstates of the coin operators or (when acted upon by these coins) such that  they transfer wave functions amplitudes onto opposite sides of the hub site. A DC cage may be seen as consisting of two ``walls", corresponding to hub sites such that  the QW wavefunction does not get transferred on the opposite states, e.g. are proportional to one of the above eigenstates. One can then analyze the QW evolution. 
 

Starting from a state localized on a site $a$, we decompose the state (after application of the local coin) into the new basis $\{\ket{R^+},\ket{R^-},\ket{L^+},\ket{L^-} \}$. The next quantum walk steps (until reaching a new hub site) consist in applying the operator: $K=S.U_2.S=K_{\mathrm{in}}+K_{\mathrm{out}}$. $S$ is the shift operator and $U_2$ the coin on the two rim sites. $K_{\mathrm{in}}$ refers to the part which is reflected back to the original hub site  and $K_{\mathrm{out}}$ to that part that will reach the neighbouring hub sites, on which we now focus.

 
 $K_{\mathrm{out}}$ action on a state at a hub site $n_0$ reads : 


$$
\ f=1/2:\ \left\{
    \begin{array}{ll}
      K_{\mathrm{out}}\ket{R^+,n_0}\propto\ket{L^-,n_0+1} \\
     K_{\mathrm{out}}\ket{R^-,n_0}\propto\ket{L^+,n_0+1} \\
    K_{\mathrm{out}}\ket{L^+,n_0}\propto\ket{R^-,n_0-1} \\
     K_{\mathrm{out}}\ket{L^-,n_0} \propto\ket{R^+,n_0-1}
    \end{array}
    \right. $$


Upon applying the $G_4$ and/or $H_4$ coins onto the QW state, a cage wall is reached whenever the part of that state on  a given site is an eigenvector of the corresponding coin. It does not generically occur after one step, but requires that a new sequence $K_{\mathrm{out}}$ is repeated, followed by $H_4$ or $G_4$, which simulates the two time-step QW operator $W^2$  for the "out" part. The Table \ref{tablesequence} (part $f=1/2$)summarizes the different sequences where cages occur. The pertinent sequences have from one to five coin operators, and their label is given in the Table.

\begin{table*}
\begin{tabular}{|c|}
\hline
 $f=0$   \\\hline \hline
  sequence: $H_4 H_4$ \\ \hline
  $\ket{R^+,n_0-1}\overset{H_4}{\rightarrow} \ket{R^+,n_0-1}\Rightarrow$cage\\
  $\ket{L^-,n_0+1}\overset{H_4}{\rightarrow} \ket{L^-,n_0+1}\Rightarrow$cage\\
    $\ket{R^-,n_0-1}\overset{H_4}{\rightarrow} \ket{L^+,n_0-1}\overset{K_{\mathrm{out}}}{\rightarrow}\ket{R^+,n_0-2}\overset{H_4}{\rightarrow}\ket{R^+,n_0-2}\Rightarrow$cage\\
    $\ket{L^+,n_0+1}\overset{H_4}{\rightarrow} \ket{R^-,n_0+1}\overset{K_{\mathrm{out}}}{\rightarrow}\ket{L^-,n_0+2}\overset{H_4}{\rightarrow}\ket{L^-,n_0+2}\Rightarrow$cage\\
    \hline
    sequence: $H_4 G_4 H_4$\\\hline
    $\ket{R^+,n_0-1}\overset{H_4}{\rightarrow} \ket{R^+,n_0-1}\Rightarrow$cage\\
  $\ket{L^-,n_0+1}\overset{H_4}{\rightarrow} \ket{L^-,n_0+1}\Rightarrow$cage\\
    $\ket{L^+,n_0+1}\overset{H_4}{\rightarrow} \ket{R^-,n_0+1}\overset{K_{\mathrm{out}}}{\rightarrow}\ket{L^-,n_0+2}\overset{G_4}{\rightarrow}\ket{L^-,n_0+2}\Rightarrow$cage\\
  $\ket{R^-,n_0-1}\overset{H_4}{\rightarrow} \ket{L^+,n_0-1}\overset{K_{\mathrm{out}}}{\rightarrow}\ket{R^+,n_0-2}\overset{G_4}{\rightarrow}\ket{L^+,n_0-2}\overset{K_{\mathrm{out}}}{\rightarrow}\ket{R^+,n_0-3}\overset{H_4}{\rightarrow}\ket{R^+,n_0-3}\Rightarrow$cage\\\hline
  sequence: $G_4 H_4 G_4 \ \mathrm{or} \ G_4 H_4 H_4$\\\hline
    $\ket{R^-,n_0-1}\overset{G_4}{\rightarrow} \ket{R^-,n_0-1}\Rightarrow$cage\\
  $\ket{L^-,n_0+1}\overset{G_4}{\rightarrow} \ket{L^-,n_0+1}\Rightarrow$cage\\
    $\ket{R^+,n_0-1}\overset{G_4}{\rightarrow} \ket{L^+,n_0-1}\overset{K_{\mathrm{out}}}{\rightarrow}\ket{R^+,n_0-2}\overset{H_4}{\rightarrow}\ket{R^+,n_0-2}\Rightarrow$cage\\
  $\ket{L^+,n_0+1}\overset{G_4}{\rightarrow} \ket{R^+,n_0+1}\overset{K_{\mathrm{out}}}{\rightarrow}\ket{L^+,n_0+2}\overset{H_4}{\rightarrow}\ket{R^-,n_0+2}\overset{K_{\mathrm{out}}}{\rightarrow}\ket{L^-,n_0+3}\overset{G_4\ \mathrm{or}\ H_4}{\rightarrow}\ket{L^-,n_0+3}\Rightarrow$cage\\\hline \hline
  $f=1/2$   \\\hline \hline
  sequence: $G_4 G_4$ \\ \hline
  $\ket{R^-,n_0-1}\overset{G_4}{\rightarrow} \ket{R^-,n_0-1}\Rightarrow$cage\\
  $\ket{L^-,n_0+1}\overset{G_4}{\rightarrow} \ket{L^-,n_0+1}\Rightarrow$cage\\
    $\ket{R^+,n_0-1}\overset{G_4}{\rightarrow} \ket{L^+,n_0-1}\overset{K_{\mathrm{out}}}{\rightarrow}\ket{R^-,n_0-2}\overset{G_4}{\rightarrow}\ket{R^-,n_0-2}\Rightarrow$cage\\
    $\ket{L^+,n_0+1}\overset{G_4}{\rightarrow} \ket{R^+,n_0+1}\overset{K_{\mathrm{out}}}{\rightarrow}\ket{L^-,n_0+2}\overset{G_4}{\rightarrow}\ket{L^-,n_0+2}\Rightarrow$cage\\
    \hline
    sequence: $H_4 G_4 H_4 \ \mathrm{or} \ H_4 G_4 G_4$\\\hline
    $\ket{R^+,n_0-1}\overset{H_4}{\rightarrow} \ket{R^+,n_0-1}\Rightarrow$cage\\
  $\ket{L^-,n_0+1}\overset{H_4}{\rightarrow} \ket{L^-,n_0+1}\Rightarrow$cage\\
    $\ket{L^+,n_0+1}\overset{H_4}{\rightarrow} \ket{R^-,n_0+1}\overset{K_{\mathrm{out}}}{\rightarrow}\ket{L^+,n_0+2}\overset{G_4}{\rightarrow}\ket{R^+,n_0+2}\overset{K_{\mathrm{out}}}{\rightarrow}\ket{L^-,n_0+3}\overset{H_4\ \mathrm{or}\ G_4}{\rightarrow}\ket{L^-,n_0+3}\Rightarrow$cage\\
  $\ket{R^-,n_0-1}\overset{H_4}{\rightarrow} \ket{L^+,n_0-1}\overset{K_{\mathrm{out}}}{\rightarrow}\ket{R^-,n_0-2}\overset{G_4}{\rightarrow}\ket{R^-,n_0-2}\Rightarrow$cage\\\hline
  sequence: $G_4 H_4 G_4 $\\\hline
    $\ket{R^-,n_0-1}\overset{G_4}{\rightarrow} \ket{R^-,n_0-1}\Rightarrow$cage\\
  $\ket{L^-,n_0+1}\overset{G_4}{\rightarrow} \ket{L^-,n_0+1}\Rightarrow$cage\\
    $\ket{R^+,n_0-1}\overset{G_4}{\rightarrow} \ket{L^+,n_0-1}\overset{K_{\mathrm{out}}}{\rightarrow}\ket{R^-,n_0-2}\overset{H_4}{\rightarrow}\ket{L^+,n_0-2}\overset{K_{\mathrm{out}}}{\rightarrow}\ket{R^-,n_0-3}\overset{G_4}{\rightarrow}\ket{R^-,n_0-3}\Rightarrow$cage\\
  $\ket{L^+,n_0+1}\overset{G_4}{\rightarrow} \ket{R^+,n_0+1}\overset{K_{\mathrm{out}}}{\rightarrow}\ket{L^-,n_0+2}\overset{H_4}{\rightarrow}\ket{L^-,n_0+2}\Rightarrow$cage\\\hline

\end{tabular}
 \caption{Effects of different sequences of coins on states $(\ket{R^+},\ket{R^-},\ket{L^+},\ket{L^-})$. All these sequences lead to a cage. $\ket{A}\overset{O}{\rightarrow}\ket{B}$ means that the application of operator $O$ onto $\ket{A}$ gives a state proportional to $\ket{B}$ }
 \label{tablesequence}
 \end{table*}

 The sequence
  $G_4 G_4$ at $f=1/2$ corresponds to the usual cages described in the paper.
  
  Given an initial site $n_0$, the caging effect should be independent of the internal state. As a consequence, since applying the first local coin only changes this internal state, the nature of the coin ($G_4$ or $ H_4$) is irrelevant, leading to the first rule.

   The right-hand side part of the QW is governed by the initial vectors $\ket{L^{+/-},n_0+1}$. $\ket{L^-}$ is an eigenvector of $H_4$ and $G_4$ so it is immediately reflected. We deduce from  Table \ref{tablesequence} that whenever $\ket{L^+}$ meets  a
   $G_4$ at
    $f=1/2$ it is  trapped on the next $a$ site. This is the second rule.

 The left-hand side part of the QW is governed by the vectors $\ket{R^{+/-},n_0-1}$.
 No matter the coin operator on the first left neighbour of the initial state, the QW will be in the state $\ket{R^-,n_0-2}$ after two time-step operations. If the next coin is $H_4$, the state is transmitted onto the next left site, still in the $\ket{R^-}$ configuration. If it is the $G_4$  coin, $\ket{R^-}$ is an eigenvector, the state is trapped. This is the third rule.

  A similar reasoning works for the $f=0$ case, upon exchanging the roles of $H_4$ and $G_4$ (see the corresponding section in Table \ref{tablesequence}).

 For completeness, if a state is initially localized on a rim site $b$ or $c$, after one time-step QW operation, the state is localized onto two hub sites $a$. Therefore, by linearity, one can apply the previous reasoning onto each of these two sites to determine the extension of the cage.

 \newpage


\providecommand{\noopsort}[1]{}\providecommand{\singleletter}[1]{#1}%

\end{document}